\documentclass{article}

\usepackage{spconf,amsmath,graphicx}
\usepackage[hidelinks]{hyperref}
\usepackage{multirow}
\usepackage{enumitem}
\usepackage{color}
\usepackage{booktabs}

\ninept

\title{Adapting Diffusion-Based Music Synthesis to Speech and Singing Voice Conversion}
\name{Ben Maman$^{\star}$,
Frank Zalkow$^{\dagger}$,
Hans-Ulrich Berendes$^{\star}$,
Paolo Sani$^{\dagger}$,
Christian Dittmar$^{\dagger}$,
Meinard Müller$^{\star}$
}
\address{$^{\star}$ International Audio Laboratories Erlangen, Germany \\
  $^{\dagger}$ Fraunhofer IIS, Erlangen, Germany}
\begin{document}
\ifpdf 
  \DeclareGraphicsExtensions{.png,.jpg,.pdf}
\else  
  \DeclareGraphicsExtensions{.eps}
\fi


\maketitle

\begin{abstract}
Recent diffusion-based generative models have achieved strong results in domain-specific audio generation tasks such as speech, singing, and instrumental music synthesis. However, these models are typically specialized and do not generalize well to mixed or intermediate audio types. In this work, we adapt a diffusion-based model originally designed for multi-instrument music synthesis to voice conversion, covering both speech and singing within a unified framework. Specifically, we extend musical note-based conditioning to include phonetic posteriorgrams (PPGs) and pitch contours, and reinterpret timbre conditioning as speaker or singer identity via feature-wise linear modulation. Experiments show that the adapted model matches or surpasses a dedicated voice conversion system in terms of naturalness and performer similarity, while maintaining accurate pitch control across speech and singing. At the same time, we observe limitations in phonetic fidelity and a degradation in vocal quality when incorporating instrumental training data. Furthermore, we demonstrate that off-the-shelf feature extractors provide effective conditioning signals, enabling large-scale self-supervised training without manual annotations. These results highlight the potential of cross-domain model transfer towards unified audio generation systems capable of handling speech, singing, and music. Qualitative samples can be found on our project page:
\url{https://benadar293.github.io/voice-conversion}
\end{abstract}

\section{Introduction}
\label{sec:intro}
Generative modeling has become a central paradigm in modern machine learning, enabling the synthesis of complex data across modalities such as images and audio. In the audio domain, recent advances have led to highly realistic generation of speech~\cite{PopovVGSKW22_DIFFVC_ICLR}, singing~\cite{LiuLRCZ22_DIFFSINGER_AIII, DaiLVG2024EXPRESSIVESINGER_ACMMULTIMEDIA, DaiWDJ2025_Everyone_ICASSP}, and instrumental music~\cite{HawthorneSRZGME22_SynthesisDiffusion_ISMIR, MamanZMB24_PerformanceConditioning_ICASSP, MamanZMB2024_MultiAspect_TASLP}, based on neural architectures that model spectral representations and convert them into waveforms.

While each of these domains, namely speech, singing, and instrumental music, has seen substantial progress through specialized models, real-world audio often combines multiple content types and attributes, including vocals, musical instruments, and hybrid or intermediate styles such as rap or Sprechgesang. However, domain-specific models do not fully capture the diversity and complexity of real-world audio, motivating the development of more unified approaches.

Among recent approaches, diffusion-based models have shown particularly strong performance in audio generation, especially if combined with structured conditioning. In particular, diffusion-based multi-instrument music synthesis~\cite{MamanZMB24_PerformanceConditioning_ICASSP, MamanZMB2024_MultiAspect_TASLP} demonstrates that a single attention-based model can generate complex musical signals with diverse instruments, timbres, and acoustic conditions. These models rely on conditioning signals such as musical score representations and timbre descriptors to control generation. Although such approaches have so far been limited to instrumental music, their design suggests that similar conditioning principles may extend to other audio domains, including speech and singing.

In this work, we investigate this hypothesis by adapting a diffusion model developed for multi-instrument music synthesis to voice conversion, covering both speech and singing within a unified framework. The key idea is to transfer and extend the conditioning mechanisms used in music synthesis: we augment note-based conditioning with phonetic posteriorgrams (PPGs) and pitch ($f_0$) contours, to capture linguistic and prosodic content, and reinterpret timbre conditioning as speaker or singer identity via feature-wise linear modulation. In this way, instrumental music synthesis and voice conversion can be formulated within a common conditional generation framework.

Our main contribution is an extensive evaluation showing that an attention-based diffusion model originally designed for instrumental music synthesis matches or surpasses a dedicated voice conversion model in naturalness and performer similarity, while maintaining accurate pitch control across speech and singing. At the same time, we identify limitations in phonetic fidelity and degradation in vocal quality when incorporating instrumental training data, highlighting important trade-offs in unified modeling.

As a second contribution, we show that off-the-shelf feature extractors provide effective conditioning signals, enabling training without manually annotated data. In particular, independently trained feature extractors can be combined to construct large-scale pseudo-annotated datasets, allowing self-supervised training across diverse domains. We further demonstrate that this approach remains effective under domain shift, such as when applying speech-trained phonetic models to singing data.

The remainder of this paper is structured as follows. Section~\ref{sec:related} reviews related work. Section~\ref{sec:method} describes the proposed method. Section~\ref{sec:experiments} presents the experimental setup and evaluation results. Section~\ref{sec:conclusion} concludes the paper.

\begin{figure*}[t!]
\centering
\includegraphics[width=0.99\textwidth]{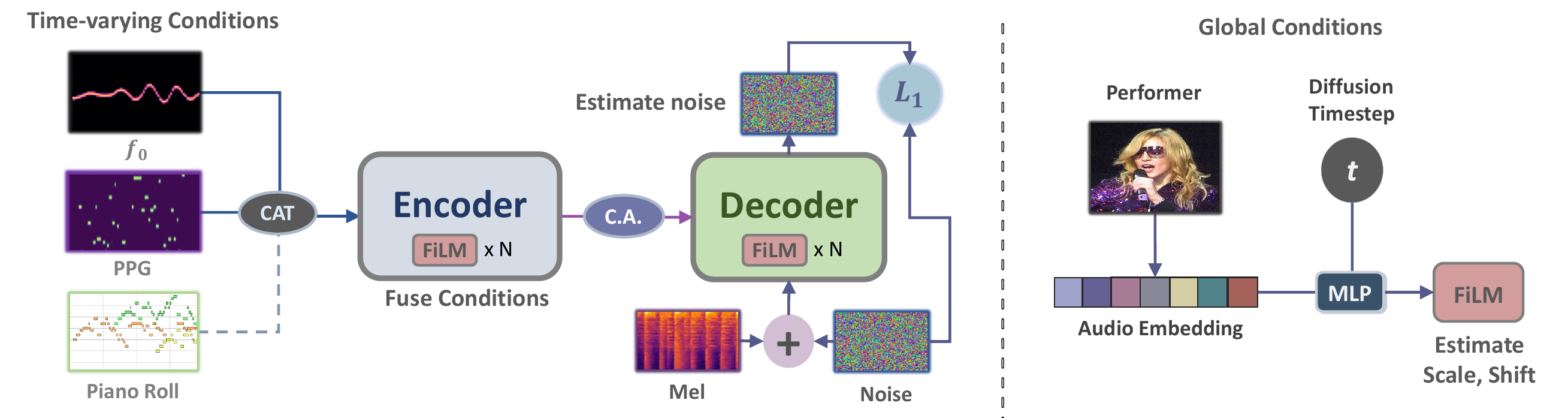}
\caption{Overview of our model. ``CAT'': Concatenation along the channel axis, ``C.A.'': Cross-attention, `N': Number of stacked blocks.}
\label{fig:overview}
\end{figure*}

\section{Related Work}\label{sec:related}
\textbf{Speech Voice Conversion (VC)} aims to transform a source speech audio sample to match the speaker identity of a target speaker~\cite{KameokaKTH2018_Stargan_SLT, SismanYKL21_VCOVERVIEW_TASLP}. High-quality conversion is achieved using various generative models, such as generative adversarial networks (GANs)~\cite{KameokaKTH2018_Stargan_SLT}, diffusion models~\cite{PopovVGSKW22_DIFFVC_ICLR}, or more recently flow-matching models~\cite{MehtaTBSH2024_Matcha_ICASSP, PiaSME2025_Flowmac_ICASSP}. To capture time-varying content from the source recording, previous work applies existing feature extractors for phonetic and pitch content, which serve as conditioning signals~\cite{KalitaDSZHP24_PADVC_IWAENC, PiaSME2025_Flowmac_ICASSP, ChurchwellMP2024_HIGH_ICASSPW}.

\textbf{Singing Voice Conversion (SVC)} is the task of adapting a source singing sample to match a target singer’s identity. The analogy between VC and SVC allows the use of shared feature representations, such as time-aligned lyrics or transcripts for linguistic content, and pitch contours for sung melody or spoken prosody. Building upon this principle,~\cite{DaiWDJ2025_Everyone_ICASSP} proposes a unified speech and singing VC model trained jointly on both domains, which enables singing VC using a speech reference as a singer-identity condition. 

Although the tasks are similar, speech benefits from abundant data, while singing synthesis suffers from severe data scarcity, as solo singing recordings are relatively rare. Some efforts have focused on creating datasets and refining existing ones~\cite{DaiWCHD23_SINGSTYLE111_ISMIR}, including lyrics and melody annotations, but their scale is still small compared to the vast amount of speech data. To address this scarcity, prior work often resorts to source-separated vocals, followed by lyrics–audio alignment~\cite{LiuLRCZ22_DIFFSINGER_AIII, DaiLVG2024EXPRESSIVESINGER_ACMMULTIMEDIA}.

\textbf{Instrumental Music Synthesis} is the task of generating musical audio from symbolic scores, often simplified using MIDI files or piano rolls. While early work focused on single instruments or monophonic music, recent approaches enable multi-instrument synthesis~\cite{HawthorneSRZGME22_SynthesisDiffusion_ISMIR} conditioned on version~\cite{MamanZMB24_PerformanceConditioning_ICASSP, MamanZMB2024_MultiAspect_TASLP}. Here, version refers to the specific recording or performer, enabling the model to capture acoustics and style. In this work, we use similar architectures. We show that version can be extended to aspects such as speaker or singer identity. Similarly, we extend conditioning on the musical score to phonetic content and vocal pitch. Together, these features provide a unified framework capable of handling instrumental music, singing, speech, or combinations thereof.

\textbf{Audio Foundation Models}~\cite{ShenJTLLHQZB24_NaturalSpeech_ICLR, Copet2023_Simple_NEURIPS} are pre-trained neural networks that learn general-purpose audio representations from diverse datasets, capturing acoustic, linguistic, and musical features across domains. They are capable of generating rich and complex audio, including singing with instrumental accompaniment. Such models are typically prompted using descriptive text or other high-level metadata, providing a form of weak conditioning. A common drawback is that conditioning is often coarse and high-level, offering limited control over fine-grained, time-varying aspects of generated audio, such as melody, phonetic content, or musical score.

\section{Method}\label{sec:method}
Following previous work in speech, singing, and instrumental music synthesis, our acoustic model is based on mel spectrogram diffusion.
To convert the generated mel spectrogram into a waveform, we use an off-the-shelf BigVGAN vocoder~\cite{LeePGCY23_BIGVGAN_ICLR}.
We choose a general purpose vocoder rather than a vocal-only one---although this may come at the expense of quality when considering vocal-only synthesis---in order to assess potential for generating vocal--instrumental 
mixtures.

\subsection{Acoustic Model Overview}
Figure~\ref{fig:overview} presents an overview of our acoustic model. Generation is factorized into three components:

\textbf{(i) Spectrogram Generation} is done using a diffusion-based spectral decoder trained to estimate noise from noisy mel spectrograms, serving as the generative backbone.

\textbf{(ii) Time-varying Conditioning} is performed via an encoder that learns a fused representation of the provided time-varying conditions, which is then supplied to the spectral decoder as auxiliary input to guide generation. We extend the piano roll used to encode musical scores~\cite{HawthorneSRZGME22_SynthesisDiffusion_ISMIR, MamanZMB2024_MultiAspect_TASLP},
with a phonetic posteriorgram (PPG) representing phonetic content, along with an $f_0$ contour representing vocal pitch. Each pitch or phonetic class is treated as a separate channel, and the features are concatenated along the channel axis.

\textbf{(iii) Global Conditioning (Performer, Diffusion Timestep)} is implemented via feature-wise linear modulations (FiLM)~\cite{PerezSVDC18_FiLM_AAAI} applied to hidden features at each block (scale, shift). These are conditioned on an audio embedding representing the vocal performer, i.e., \textit{speaker or singer}. While this mechanism was used in instrumental synthesis to condition on timbre and acoustics~\cite{MamanZMB24_PerformanceConditioning_ICASSP, MamanZMB2024_MultiAspect_TASLP}, we show it is similarly effective for conditioning on the performer.


\subsection{Condition Dropout} Each condition $c$ is randomly and independently dropped with probability $p^\mathrm{drop}_c$ by zeroing out the condition. Additionally, all conditions are jointly dropped with probability $p^\mathrm{drop}_\mathrm{all}$ to ensure unconditional samples occur sufficiently often during training. Condition dropout promotes robustness and enables generation from partial conditions, such as a speech sample conditioned on phonetic content (PPG) alone, with other aspects like the pitch contour generated freely. It is also required for classifier-free guidance~\cite{HoS22_ClassifierFreeGuidance_arXiv}, enabling control over conditioning strength.

\subsection{Pitch Range Adaptation}\label{sec:pitch_range}
Singers have distinct pitch ranges; therefore, transferred pitch contours should match the target singer’s natural range. For example, na{\"i}vely transferring a song from a female to a male singer may produce high-pitched, female-like singing, even when the performer condition is applied, since female voices are generally higher.

To address this, we apply pitch range adaptation, as in prior work on speech and singing voice conversion~\cite{KalitaDSZHP24_PADVC_IWAENC, DaiWDJ2025_Everyone_ICASSP}. We denote the source pitch contour in the log-frequency semitone scale as $\{p^\mathrm{src}_t\}_{t=1}^T$, where $f_t^\mathrm{src}$ denotes the frame-wise fundamental frequency and $p_t^\mathrm{src} = 12\log_2(f_t^\mathrm{src}/440) + 69$. We then perform mean shifting to adapt the source to the target range:
\begin{equation}\label{eq:pitch_range_adaptation}
    p^\mathrm{tar}_t = p^\mathrm{src}_t - \mu^\mathrm{src}_p + \mu^\mathrm{tar}_p,
\end{equation}
where $\mu_p^\mathrm{src}$ and $\mu_p^\mathrm{tar}$ denote the mean values of the source and target contours. We find averaging in the semitone scale more suitable for musical contexts than averaging in linear frequency (Hz).

\subsection{Feature Extraction}

We estimate $f_0$ using CREPE~\cite{KimSLB18_CREPE_ICASSP},
and phonetic posteriorgrams (PPGs) using a variant of wav2vec~2.0~\cite{BaevskiZMA20_WAV2VEC2_NEURIPS, BabuWTLXGSPSPBC22_XLSR_INTERSPEECH}.\footnote{\url{https://huggingface.co/vitouphy/wav2vec2-xls-r-300m-timit-phoneme}} Although the PPG extractor is trained on speech data, we demonstrate that it provides meaningful conditioning for singing signals as well. To extract vocal $f_0$ and PPG features from vocal--instrumental mixtures, we first apply a vocal source separation model~\cite{RouardMD23_HTDemucs_ICASSP}.
For the joint vocal--instrumental modeling setting, we additionally estimate piano rolls using a multi-instrument extension of Onsets and Frames~\cite{HawthorneESRSRE18_OnsetsFrames_ISMIR, HawthorneSRSHDE19_MAESTRO_ICLR, MamanBermano22_UnalignedAMT_ICML, YaffeMMB25_CountNotes_ISMIR}.
Finally, to condition on performer, we extract audio embeddings using TRILL~\cite{ShorJMLTQTSEH20_NonSemanticSpeechRepresentation_Interspeech},
which is trained with a triplet loss objective such that temporally proximal audio segments are mapped closer together in the embedding space.

\subsection{Architecture}\label{sec:architecture}
We experiment with two state-of-the-art models, both based on publicly available implementations. The first model, developed for instrumental music synthesis, is a T5-based diffusion model~\cite{HawthorneSRZGME22_SynthesisDiffusion_ISMIR, MamanZMB24_PerformanceConditioning_ICASSP, MamanZMB2024_MultiAspect_TASLP}. The second model, developed for voice conversion, is based on FlowMAC~\cite{PiaSME2025_Flowmac_ICASSP}, which builds on MatchaTTS~\cite{MehtaTBSH2024_Matcha_ICASSP}. It is conditioned using the the PAD-VC voice conversion system~\cite{KalitaDSZHP24_PADVC_IWAENC}, based on Forward Tacotron,\footnote{\url{https://github.com/axelspringer/ForwardTacotron}} as done in~\cite{ZalkowSKHPD25_LowResourceGenerativePostprocessing_INTERSPEECH}.


We choose the music synthesis model for its ability to generate complex, multi-instrument signals under varied acoustic conditions, suggesting potential for singing with accompaniment. The speech model serves as a vocal-specialized baseline, allowing comparison and evaluation of the music model’s ability to handle vocal and phonetic content. While many existing diffusion-based singing voice conversion systems lack publicly available implementations, potentially limiting reproducibility, this does not significantly hinder our evaluation; meaningful benchmarks for both speech and singing can still be obtained from speech VC systems, as singing and speech VC systems share largely analogous principles and architectural designs, such as 1D convolution--attention U-Net hybrids previously employed in both singing and music synthesis~\cite{DaiLVG2024EXPRESSIVESINGER_ACMMULTIMEDIA, DaiWDJ2025_Everyone_ICASSP, MamanZMB24_PerformanceConditioning_ICASSP, MamanZMB2024_MultiAspect_TASLP}. At the same time, task-specific architectural optimization remains an important direction for future research.

\textbf{Music Synthesis Model.}  
This is a diffusion model based on the T5 Transformer architecture~\cite{HawthorneSRZGME22_SynthesisDiffusion_ISMIR, MamanZMB24_PerformanceConditioning_ICASSP, MamanZMB2024_MultiAspect_TASLP}, comprising a conditioning encoder and a spectral decoder, each built from stacks of self-attention layers. The decoder receives noisy spectrograms and predicts the added noise, while also incorporating time-varying conditioning from the encoder via interleaved cross-attention. Performer conditioning (singer or speaker) is applied to both encoder and decoder through FiLM layers, whereas noise-level conditioning is applied only in the decoder. The model is trained end-to-end using an $L_1$ loss on the predicted noise.

\textbf{Voice Conversion Model.}  
This model, FlowMAC~\cite{PiaSME2025_Flowmac_ICASSP}, is a flow-matching model that combines 1D convolutional residual blocks with attention layers. It is based on the MatchaTTS~\cite{MehtaTBSH2024_Matcha_ICASSP} decoder, omitting the text encoder since the conditioning PPGs and $f_0$ contours are time-aligned with the audio. The model is conditioned on the output of PAD-VC (described below). Drawing an analogy to the T5 model, the PAD-VC conditioning component corresponds to the T5 time-varying conditioning encoder, while the FlowMAC decoder corresponds to the T5 spectral decoder.

\textbf{PAD-VC~\cite{KalitaDSZHP24_PADVC_IWAENC}.}  
The model consists of 1D convolutional blocks combined with long short-term memory (LSTM) layers. It is based on the ForwardTacotron decoder, omitting the text encoder since the PPG and $f_0$ are time-aligned (as above).
It is trained using a spectral $L_1$ reconstruction loss. Although its output provides only a coarse spectrogram estimate of relatively low quality, it primarily serves as a conditioning signal for the FlowMac model. In addition, we use this coarse output as a lower anchor for evaluation.

For all models, training and sampling procedures follow the original publications.

\section{Experiments}\label{sec:experiments}
\subsection{Datasets}\label{sec:datasets}
We use a large compound dataset of unannotated audio from different domains, including speech, singing, instrumental music, and mixed recordings. While these datasets are internal, all feature extractors used to automatically generate pseudo-annotations are publicly available, allowing replication of our approach with equivalent audio data. The specific datasets used are as follows:

(i) \textbf{Speech ($\mathcal{D}_{\mathrm{speech}}$):} A $\sim$33\,h compound dataset comprising proprietary and public sources, featuring recordings in the English language from five speakers---two male
and three female. As a held-out set, we randomly sample 50 full-utterance excerpts from each of the five speakers, each excerpt 2--12\,s long, totaling $\sim$19\,m, yielding a split $\mathcal{D}_{\mathrm{speech}}=\mathcal{D}_{\mathrm{speech}}^{\mathrm{train}}\cup\mathcal{D}_{\mathrm{speech}}^{\mathrm{test}}$.

(ii) \textbf{Singing ($\mathcal{D}_{\mathrm{sing}}$):} A $\sim$31\,h compound dataset consisting of the following: The SingStyle111 dataset~\cite{DaiWCHD23_SINGSTYLE111_ISMIR} which contains $\sim$13\,h of solo singing from four male and four female singers, together with the source-separated vocals from the following $\mathcal{D}_{\mathrm{mix}}$ which includes four female and 33 male singers, using only voice-active regions totaling $\sim$18\,h. Source separation was verified to be of high quality through informal listening tests, to ensure it does not affect evaluation. As a held-out set, we randomly sample three songs for each of the eight singers in SingStyle111, each song 1--7\,m long,
totaling $\sim$1.3\,h,
yielding a split: $\mathcal{D}_{\mathrm{sing}}=\mathcal{D}_{\mathrm{sing}}^{\mathrm{train}}\cup\mathcal{D}_{\mathrm{sing}}^{\mathrm{test}}$.

 (iii) \textbf{Vocal--Instrumental Mix ($\mathcal{D}_{\mathrm{mix}}^{\mathrm{train}}$):} A $\sim$90\,h compound dataset including the Schubert Winterreise dataset~\cite{WeissZAMKVG21_WinterreiseDataset_ACM-JOCCH} ($\sim$11\,h, nine male singers with piano accompaniment), pop and rock music ($\sim$34\,h, 24 male and four female singers) and instrumental Western classical music ($\sim$47\,h). For evaluation, we use a $\sim$3\,h test set $\mathcal{D}_{\mathrm{mix}}^{\mathrm{test}}$ comprising 40 well-known popular songs not included in $\mathcal{D}_{\mathrm{mix}}^{\mathrm{train}}$, from 14 singers whose other songs are in $\mathcal{D}_{\mathrm{mix}}^{\mathrm{train}}$ (11 male, 3 female), and 6 singers not in $\mathcal{D}_{\mathrm{mix}}^{\mathrm{train}}$ (2 male, 4 female).

 Lastly, we denote by $\mathcal{D}_{\mathrm{voc}}$ the union of the vocal training data:
\begin{equation}
    \mathcal{D}_{\mathrm{voc}} =
    \mathcal{D}_{\mathrm{speech}}^{\mathrm{train}} \cup
    \mathcal{D}_{\mathrm{sing}}^{\mathrm{train}},
\end{equation}
and by $\mathcal{D}_{\mathrm{all}}$ the union of all three training sets:
\begin{equation}
    \mathcal{D}_{\mathrm{all}} =
    \mathcal{D}_{\mathrm{voc}} \cup
    \mathcal{D}_{\mathrm{mix}}^{\mathrm{train}} =
    \mathcal{D}_{\mathrm{speech}}^{\mathrm{train}} \cup
    \mathcal{D}_{\mathrm{sing}}^{\mathrm{train}} \cup
    \mathcal{D}_{\mathrm{mix}}^{\mathrm{train}}.
\end{equation}

\subsection{Evaluation}\label{sec:speech_singing_vc}
\begin{figure}[t!]
\centering
\includegraphics[width=0.85\columnwidth]{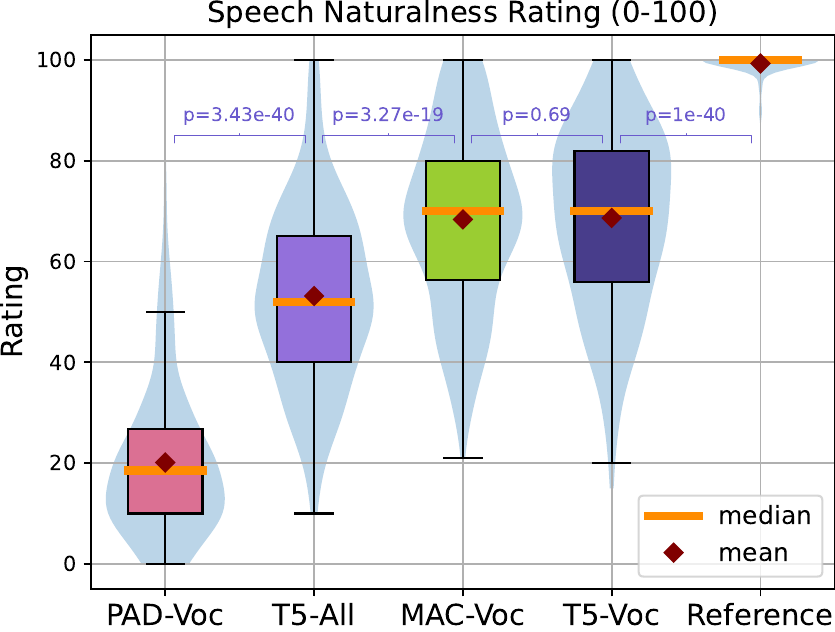}
\caption{Speech naturalness listening test results.}
\label{fig:mushra_speech}
\end{figure}
\begin{figure}[t!]
\centering
\includegraphics[width=0.85\columnwidth]{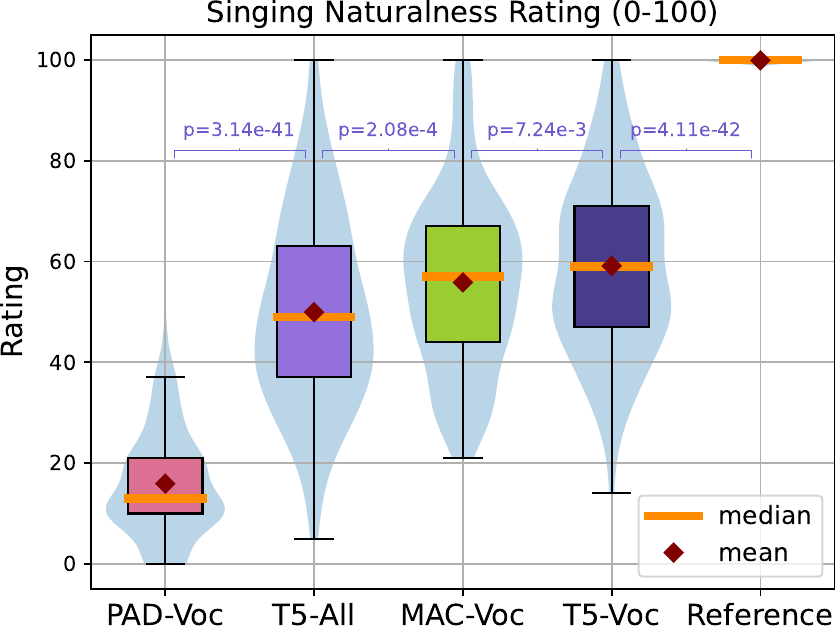}
\caption{Singing naturalness listening test results.}
\label{fig:mushra_singing}
\vspace{-0.1cm}
\end{figure}

Audio synthesizers, whether for speech, singing, or music, should meet several requirements, some of which may be interdependent:

(i) \textbf{Quality and naturalness:} Generated audio should be natural, high-quality, and ideally indistinguishable from real recordings, with realistic intonation, prosody, and expressive nuances.

(ii) \textbf{Performer and acoustic control:} The system should allow control over speaker, singer, or performer identity, including timbre, style, and acoustic environment.

(iii) \textbf{Pitch, prosody, and musical control:} The system should accurately reproduce pitch contours, prosody, and musical structure, including melody, harmony, timing, and instrumentation. 

(iv) \textbf{Linguistic and phonetic intelligibility and control:} Generated speech and singing should be intelligible, preserving the source linguistic, phonetic, and lyrical content.




In our evaluation, we aim to cover these requirements as comprehensively as possible. We conduct both qualitative evaluation through listening tests and quantitative evaluation using objective metrics. The listening tests focus on the two previously mentioned aspects: (i) quality and naturalness (Section~\ref{sec:naturalness_listening_test}), and (ii) control over performer identity (Section~\ref{sec:similarity}). We complement these tests with a quantitative evaluation of the same aspects using the Fr{\'{e}}chet Audio Distance (FAD)~\cite{KilgourZRS19_FAD_Interspeech} (Sections~\ref{sec:fad} and \ref{sec:mixed}).

In addition,
we quantitatively evaluate
the faithfulness of the generated signals in the remaining two aspects: (iii) pitch control, i.e., how faithfully the original pitch contour ($f_0$) is maintained (Section~\ref{sec:f0}); and (iv) linguistic and phonetic fidelity, i.e., how accurately the phonetic content is preserved (Section~\ref{sec:phonetic_control}).

We compare the following three models trained on the vocal data $\mathcal{D}_\mathrm{voc}$: \texttt{T5-Voc}, a T5 diffusion-based model; \texttt{PAD-Voc}, a ForwardTacotron-based model trained with a spectral reconstruction loss; and \texttt{MAC-Voc}, a FlowMAC model conditioned on the output of \texttt{PAD-Voc}.

To assess the T5 model's ability to handle mixed data, we include \texttt{T5-All}, trained on $\mathcal{D}_\mathrm{all}$, excluding the held-out sets. For this model, we condition on the instrumental musical score using a piano roll~\cite{MamanZMB2024_MultiAspect_TASLP}, concatenated with the vocal conditioning features ($f_0$ and PPG). To better handle the different data types within this single model, we further extend its conditioning by concatenating a one-hot encoding of the data type (vocal-only or vocal with instrumentals) to the performer condition.

\subsubsection{Naturalness Listening Tests}\label{sec:naturalness_listening_test}
\begin{table}[t!]
\begin{center}
\begin{tabular}{ccc}
\toprule
\multirow{2}{*}{} & \multicolumn{2}{c}{Naturalness Rating$\uparrow$} \\

\cmidrule{2-3}
& Speech & Singing \\
\midrule
\texttt{T5-Voc} & \textbf{68.63$\pm$18.11} & \textbf{59.11$\pm$17.54} \\
\texttt{T5-All} & 53.15$\pm$17.66 & 49.94$\pm$19.27 \\
\texttt{PAD-Voc} & 20.12$\pm$15.55 & 15.90$\pm$10.57 \\
\texttt{MAC-Voc} & 68.34$\pm$17.98 & 55.84$\pm$17.53 \\
\hline
\texttt{Ref.} & \hspace{-5pt}99.28$\pm$4.49 & \hspace{-5pt} 99.89$\pm$1.64 \\


\bottomrule
\end{tabular}
\caption{Speech and singing naturalness listening tests results.}
\label{table:mushra}
\end{center}
\vspace{-0.5cm}
\end{table}
To evaluate the quality of the synthesized audio in terms of naturalness and realism, we follow a listening test protocol with a hidden reference and a lower anchor similar to a standard Multiple Stimuli with Hidden Reference and Anchor (MUSHRA) test~\cite{ITU15}, which often enables statistically significant comparisons with a relatively small number of participants. We perform two listening tests, one for speech and one for singing, comparing the four aforementioned models, in the task of reconstructing the original audio from the conditioning features, namely $f_0$, PPG, and performer identity. In this setting, the target identity is the same as in the source audio excerpt from which the $f_0$ and PPG were extracted.

As a reference we use the vocoded version of the original audio
(i.e., reconstructed by the vocoder from the mel spectrogram),
which is the upper bound on the attainable quality. Listeners who rated the hidden reference lower than 95 more than once were discarded (post-screening). In the speech listening test, 26 listeners participated, one of which was post-screened, leaving 25 listeners. In the singing listening test, 29 listeners participated, four of which were post-screened, leaving 25 listeners.

Results for speech and singing appear in box plot and violin plot form in Figures~\ref{fig:mushra_speech} and~\ref{fig:mushra_singing}, with pairwise $p$-values using a Wilcoxon signed-rank test. Mean and standard deviation values appear in Table~\ref{table:mushra}. It can be seen that \texttt{T5-Voc} performs comparably or slightly better than \texttt{MAC-Voc}. \texttt{T5-Voc} was rated slightly higher than \texttt{MAC-Voc} in singing---with a mean rating of $59.11$ compared to $55.84$, and a $p$-value of $7.24e^{-3}$. In speech, the difference (68.63 compared to 68.34) is not significant ($p=0.69$). We observe generally lower ratings for singing, which may indicate that expressive singing is harder to generate than speech. Finally, the \texttt{PAD-Voc} model achieves scores of about $16$--$20$, indicating that a reconstruction objective alone is insufficient for this task, which requires generative modeling. Qualitative examples from both listening tests are provided on the project page.

A key takeaway from this experiment is that \texttt{T5-Voc}, adapted from instrumental music synthesis, can be effectively applied to vocal synthesis, with performance comparable to that of a specialized VC model. However, incorporating instrumental data during training (\texttt{T5-All}) leads to degradation in perceived quality for both speech and singing, by approximately 10--15 points. This outcome is not unexpected: introducing data from a distinct domain increases task complexity and may demand greater model capacity, and it is commonly observed that domain-specialized models outperform unified, all-in-one approaches. Nevertheless, a vocal-only model cannot generate vocal--instrumental mixtures, making \texttt{T5-All} necessary for such tasks. Addressing this trade-off remains an important direction for future work.

\subsubsection{Singer Similarity Listening Test}\label{sec:similarity}

In this test, we investigate how closely the generated audio resembles recordings of a target singer, comparing \texttt{T5-Voc} and \texttt{MAC-Voc} from the previous subsection in singing voice conversion. This evaluation serves two purposes: first, to examine, within the same model, how generating the same content conditioned on two different singers affects output; and second, to compare the strength of the conditioning effect across models.

For each question we randomly sample a reference singer \texttt{ref}, and another singer \texttt{other} of the same gender. A random source excerpt provides the $f_0$ and the PPG. Using each of the \texttt{T5-Voc} and \texttt{MAC-Voc} models, we sonify the source $f_0$ and PPG conditioned on \texttt{ref} and \texttt{other}, yielding four generated samples. Listeners are then presented with three random excerpts of \texttt{ref} to familiarize themselves with the target voice, and are asked to rate the similarity of each generated sample to \texttt{ref}, according to the following Likert scale: (1) ``completely different person,'' (2) ``probably different person,'' (3) ``similar,'' (4) ``probably the same person,'' and (5) ``exactly the same person.'' We find this rating scale more meaningful than the 0–100 continuous scale used in previous tests, as the labeled categories provide clearer, interpretable reference points for listeners. To isolate the influence of pitch range and performer condition, we apply pitch range adaptation using the target performer’s range, ensuring that all samples---including those conditioned on \texttt{other}---are generated at the same pitch.

The test comprised ten randomly sampled questions, each with four generated and three reference samples. Excerpts were full utterances, 3--12\,s long, drawn from a test set balanced between male and female singers. Twenty listeners participated.

\begin{figure}[t!]
\centering
\includegraphics[width=0.85\columnwidth]{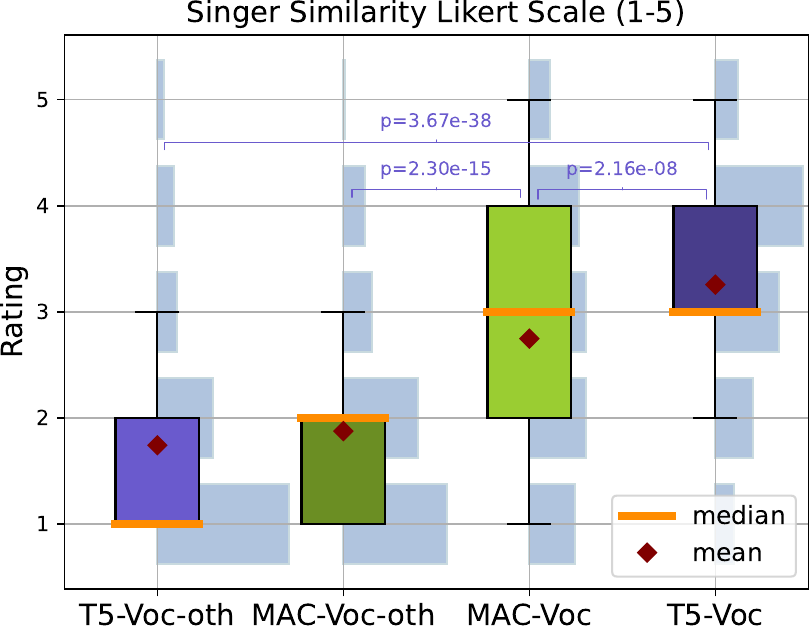}
\caption{Singer similarity listening test results.}
\label{fig:singing_similarity}
\vspace{-0.1cm}
\end{figure}

Results appear in Figure~\ref{fig:singing_similarity}, including $p$-values, and rating histograms for each Likert value and model (light-colored bars). For both \texttt{T5-Voc} and \texttt{MAC-Voc}, conditioning on the target singer substantially increases perceived similarity. For instance, \texttt{T5-Voc} ratings rise on average from $1.76\pm 1.06$ when conditioned on another singer (\texttt{T5-Voc-oth}) to $3.25\pm 1.09$ when conditioned on the reference singer (\texttt{T5-Voc}), with the most frequent rating being (4) ``probably the same person''.

We observe that
\texttt{T5-Voc} achieves higher similarity ratings than \texttt{MAC-Voc} with a mean rating of $3.25$ versus $2.75$ and a $p$-value below $10^{-7}$, indicating stronger conditioning. We hypothesize this stems from the conditioning in \texttt{T5-Voc} being applied both in the encoder and decoder, whereas \texttt{MAC-Voc} conditioning is done only in its ForwardTacotron-based encoder.
Conditioning its decoder could potentially improve performer similarity.

A key takeaway from this experiment is that the same conditioning technique used for acoustics in instrumental music synthesis~\cite{MamanZMB2024_MultiAspect_TASLP} is similarly effective for conditioning on the singer in human VC. Qualitative examples are available on the project page.

\subsubsection{Fr{\'{e}}chet Audio Distance (FAD)}\label{sec:fad}
We complement the listening tests with a quantitative evaluation using the Fr{\'{e}}chet Audio Distance (FAD)~\cite{KilgourZRS19_FAD_Interspeech} which measures a distributional distance between two sets. Following~\cite{HawthorneSRZGME22_SynthesisDiffusion_ISMIR, MamanZMB24_PerformanceConditioning_ICASSP, MamanZMB2024_MultiAspect_TASLP}, we use the TRILL model~\cite{ShorJMLTQTSEH20_NonSemanticSpeechRepresentation_Interspeech}, producing five audio embeddings per second. We apply two variants: \textbf{(i) All-FAD} measures overall quality by computing the distance between all generated audio and the reference set of real recordings; \textbf{(ii) Performer-FAD} measures performer similarity by computing the distance between audio generated with a specific performer condition and the set of real recordings of the same performer. To evaluate conversion quality, we render the held-out sets with randomly sampled other target performers---speakers for speech and singers for singing.

Table~\ref{table:fad} reports the results. For All-FAD, we compare generated audio to the train and test sets (cols. 2--3, 5--6). For Performer-FAD, we compare synthesized and real subsets corresponding to performers (cols. 4, 7). \texttt{T5-Voc} produces best overall scores, usually surpassing \texttt{MAC-Voc}. The difference is most prominent for Performer-FAD---for example, in singing, \texttt{T5-Voc} yields 0.141 compared to 0.178 of \texttt{MAC-Voc}, consistent with the similarity listening test, where \texttt{T5-Voc} scored higher.

We hypothesize the lower distances w.r.t. the train sets arise because the train sets are larger, thus of a smoother distribution.

While the \texttt{Vocoded} source yields best All-FAD scores, it yields worst Performer-FAD scores, since the source and target performers differ, confirming that performer conditioning shifts the generated audio distribution towards the target performer. 

It is also evident that adding instrumental data (\texttt{T5-All}) degrades scores, consistent with the listening tests (Section~\ref{sec:naturalness_listening_test}).

\begin{table}[t!]
\begin{center}
\setlength{\tabcolsep}{4pt}
\begin{tabular}{ccccccc}
\toprule
\multirow{2}{*}{} & \multicolumn{6}{c}{FAD$\downarrow$} \\
\cmidrule{2-7}
& \multicolumn{3}{c}{Speech} & \multicolumn{3}{c}{Singing} \\
\cmidrule{2-7}
& $\mathcal{D}_{\mathrm{speech}}^{\mathrm{test}}$ & $\mathcal{D}_{\mathrm{speech}}^{\mathrm{train}}$ & Perf. & $\mathcal{D}_{\mathrm{sing}}^{\mathrm{test}}$ & $\mathcal{D}_{\mathrm{sing}}^{\mathrm{train}}$ & Perf. \\
\midrule
\texttt{T5-Voc} & \textbf{0.162} & 0.118 &\textbf{0.156}& \textbf{0.108} & 0.093 & \textbf{0.141} \\
\texttt{T5-All} & 0.187 & 0.141 &0.190& 0.142 & 0.128 &0.192 \\
\texttt{PAD-Voc} & 0.345 & 0.288 &0.343& 0.271 & 0.257 & 0.356\\
\texttt{MAC-Voc} & 0.171 & \textbf{0.113} &0.163& 0.110 & \textbf{0.092} & 0.178\\
\hline
\texttt{Vocoded} & 0.002 & 0.054 & 0.573 & 0.004 & 0.065 & 0.369 \\
\bottomrule
\end{tabular}
\caption{FAD results for speech and singing.}\label{table:fad}
\end{center}
\vspace{-0.4cm}
\end{table}

\subsubsection{Vocal Pitch Control}\label{sec:f0}
\begin{table}[t!]
\begin{center}
\setlength{\tabcolsep}{4pt}
\begin{tabular}{ccccccc}
\toprule
\multirow{4}{*}{} & \multicolumn{6}{c}{$f_0$ Accuracy $\uparrow$} \\
\cmidrule{2-7}
 & \multicolumn{3}{c}{Speech} & \multicolumn{3}{c}{Singing} \\
\cmidrule{2-7}
& RPA & Ov50 & Ov25 & RPA & Ov50 & Ov25 \\
\midrule
\texttt{T5-Voc} & \textbf{83.0} & \textbf{81.9} & \textbf{71.6} & \textbf{94.7} & \textbf{92.5} & \textbf{83.1} \\
\texttt{T5-All} & 80.4 & 80.8 & 70.4 & 94.2 & 92.1 & 82.2 \\
\texttt{PAD-Voc} & 78.2 & 79.1 & 68.5 & 92.7 & 89.4 & 78.0 \\
\texttt{MAC-Voc} & 79.8 & 80.1 & 69.6 & 93.6 & 91.8 & 81.8 \\
\bottomrule
\end{tabular}
\caption{Pitch estimation results, including raw pitch accuracy at a 50-cent threshold (RPA), overall accuracy at a 50-cent threshold (Ov50), and overall accuracy at a 25-cent threshold (Ov25).}\label{table:f0}
\end{center}
\vspace{-0.4cm}
\end{table}

To evaluate vocal pitch control, i.e., how well the generated audio preserves the source pitch contour, we extract pitch contours and voice activity from both signals using CREPE and compare them using standard pitch estimation metrics~\cite{PolinerEEGSO07_MelodyTranscription_TASLP, SalamonG12_MelodyExtraction_TASLP}. We perform the evaluation on the same held-out sets with randomly sampled target performers from Section~\ref{sec:fad}. The reference contours used for evaluation are transposed versions of the source pitch contours according to the pitch range shift.

We report the \textit{raw pitch accuracy} (RPA) metric, measuring the proportion of voiced frames in which the estimated frequency is within a threshold of the reference frequency (typically 50 cents, i.e., a half semitone, or smaller), and the \textit{overall accuracy} metric, which also takes voice activity into account. The metrics are computed using \textit{mir\_eval}~\cite{RaffelMHSNLE14_MirEval_ISMIR}. While pitch estimation may introduce errors affecting evaluation, CREPE is shown to achieve over 95$\%$ accuracy across diverse settings~\cite{KimSLB18_CREPE_ICASSP}.


Table~\ref{table:f0} reports results for speech and singing using RPA (without voicing, 50 cents) and overall accuracy (with voicing, 50 and 25 cents). \texttt{T5-Voc} achieves the highest scores in most cases, although differences from \texttt{MAC-Voc} are typically 1--3\%. Singing RPA is high (92--95\%) across all models. Adding instrumental data (\texttt{T5-All}) has minimal effect, particularly for singing, with differences below 0.5\%. Notably, singing scores substantially exceed speech scores, possibly due to less stable pitch estimation on speech---an issue that requires further investigation.

In summary, the results in Table~\ref{table:f0} indicate that both models, \texttt{T5-Voc} and \texttt{MAC-Voc}, although originating from different domains (instrumental music and speech), robustly maintain pitch contours in human voice conversion.

\subsubsection{Phonetic Control}\label{sec:phonetic_control}
\begin{table}[t!]
\begin{center}
\setlength{\tabcolsep}{4pt}
\begin{tabular}{cccccc}
\toprule
\multirow{2}{*}{} & \multicolumn{4}{c}{PPG Distance $\downarrow$} \\
\cmidrule{2-5}
& \multicolumn{2}{c}{Speech} & \multicolumn{2}{c}{Singing} \\
\cmidrule{2-5}
& J-S & Wass. & J-S & Wass. \\
\midrule
\texttt{T5-Voc} & 0.338 & 4.495 & 0.266 & 2.319 \\
\texttt{T5-All} & 0.352 & 4.628 & 0.273 & 2.346 \\
\texttt{PAD-Voc} & 0.351 & 4.660 & 0.276 & 2.329 \\
\texttt{MAC-Voc} & \textbf{0.322} & \textbf{4.334} & \textbf{0.245} & \textbf{2.151} \\
\bottomrule
\end{tabular}
\caption{PPG distance results using the Jensen-Shannon divergence and the Wasserstein distance.}\label{table:ppg}
\end{center}
\vspace{-0.4cm}
\end{table}

To evaluate phonetic control, that is, how well phonetics are preserved relative to the source, we extract PPGs using wav2vec 2.0 (Section~\ref{sec:method}) and compare the source and generated audio using two posteriorgram distance metrics, based on the Jensen--Shannon divergence~\cite{FugledeT2004_Jensen_ISIT, ChurchwellMP2024_HIGH_ICASSPW} and the Wasserstein distance~\cite{Vaserstein1969_Markov}, respectively. These metrics serve as a quantitative measure of intelligibility. We perform the evaluation on the same held-out sets with randomly sampled target performers from Section~\ref{sec:fad} and Section~\ref{sec:f0}.

Table~\ref{table:ppg} reports the results. \texttt{MAC-Voc} consistently achieves best scores, whether using the Jensen--Shannon divergence or the Wasserstein distance. While \texttt{T5-Voc} produces comparable or better results than \texttt{MAC-Voc} in qualitative listening tests on naturalness, singer similarity, and pitch control (Figures~\ref{fig:mushra_speech},~\ref{fig:mushra_singing},~\ref{fig:singing_similarity} and Tables~\ref{table:mushra},~\ref{table:f0}), phonetics are better preserved by the \texttt{MAC-Voc} model according to the metrics.

We hypothesize that \texttt{T5-Voc} introduces more diversity when rendering PPGs, owing to its higher expressivity and its purely attention-based architecture, which lacks convolutional constraints. Its expressiveness is evident in qualitative examples; for instance, we have noticed that \texttt{T5-Voc} converted a German guttural `r' (i.e., a throaty `r')  to an English alveolar approximant `r'. These results indicate that high perceptual quality and generative capability do not necessarily align with phonetic accuracy, whose importance depends on the application; in some cases, generative freedom may be acceptable or even desirable. Whether the source phonetics should be preserved or, conversely, adapted to the phonetic style of the target performer is a matter of task definition and of whether such adaptation is considered desirable.

We find that incorporating instrumental data (\texttt{T5-All}) reduces phonetic metric scores, consistent with earlier experiments.

Finally, while we report quantitative metrics for phonetic fidelity and intelligibility, further evaluation through qualitative listening tests remains an important direction for future work.

\subsubsection{Vocals with Instrumentals}\label{sec:mixed}
While we have seen that the joint \texttt{T5-All} model performs worse on vocals than the vocal-specialized \texttt{T5-Voc}, in this section we present an initial experiment highlighting the strengths of \texttt{T5-All} and the benefits of the joint approach. Specifically, we conduct a basic experiment in the scenario of acoustic conversion or style transfer of singing with accompaniment, which can be viewed as a generalization of voice conversion. 

In this setting, we take a source music excerpt and analyze its components using our feature extraction pipeline (Section~\ref{sec:method}), then resynthesize the audio with a new performer condition, adjusting the pitch range (Section~\ref{sec:pitch_range}). Feature extraction includes vocal and instrumental elements as well as performer and acoustics.

We compare two setups: (i) using an instrumental piano roll only~\cite{MamanZMB2024_MultiAspect_TASLP}, and (ii) combining vocal features ($f_0$ and PPG) with the piano roll. We evaluate the results using the All-FAD metric, which serves as an indicator of fidelity to the original and, more generally, to real performances.

Results are reported in Table~\ref{table:all_fad_mixed}. We observe that the combined approach (\texttt{MIDI+F0+PPG}) yields substantial improvements, from 0.277 to 0.236 with respect to the test set and from 0.25 to 0.213 with respect to the training set, over synthesis using piano rolls alone (\texttt{MIDI}). This highlights that incorporating both vocal and instrumental features brings the generated audio closer to the original signal and increases its realism.

\begin{table}[t!]
\begin{center}
\setlength{\tabcolsep}{6pt}
\begin{tabular}{lcc}
\toprule
\multirow{2}{*}{} & \multicolumn{2}{c}{FAD $\downarrow$ (Mix)} \\
\cmidrule{2-3}
 & $\mathcal{D}_{\mathrm{mix}}^{\mathrm{test}}$ & $\mathcal{D}_{\mathrm{mix}}^{\mathrm{train}}$ \\
\midrule
\texttt{MIDI} & 0.277 & 0.250 \\
\texttt{MIDI+F0+PPG} & 0.236 & 0.213 \\
\bottomrule
\end{tabular}
\caption{FAD results on mixed vocal--instrumental data.}
\label{table:all_fad_mixed}
\vspace{-0.4cm}
\end{center}
\end{table}





\section{Conclusion}\label{sec:conclusion}
In this work we evaluated the performance of an attention-based diffusion model adapted from instrumental music synthesis to human voice conversion. Through an extensive evaluation across speech and singing we have shown it can match---or even surpass---a dedicated voice conversion model in
terms of quality and performer similarity. While our results also indicate a decline in vocal synthesis quality when including training data with instrumental music, the model's demonstrated ability to handle both instrumental and vocal synthesis within a unified framework underscores its potential for generating complex signals, such as singing with instrumental accompaniment.
Exploring and evaluating generation of such signals
remains an important direction for future work.

\section{Acknowledgments}
This work was funded by the Deutsche Forschungsgemeinschaft (DFG, German Research Foundation) under Grant No. 500643750 (MU 2686/15-1). The International Audio Laboratories Erlangen are a joint institution of
the Friedrich-Alexander-Universität Erlangen-Nürnberg (FAU) and
Fraunhofer Institute for Integrated Circuits IIS. This work was supported by Fraunhofer-Zukunfts\-stiftung. The authors gratefully acknowledge the scientific support and HPC resources provided by the Erlangen National High Performance Computing Center (NHR@FAU) of 
the Friedrich-Alexander-Universität Erlangen-Nürnberg (FAU)
under the NHR project b265dc. NHR funding is provided by federal and Bavarian state authorities. NHR@FAU hardware is partially funded by the German Research Foundation (DFG)---440719683.




\begin{thebibliography}{10}
\providecommand{\url}[1]{#1}
\csname url@samestyle\endcsname
\providecommand{\newblock}{\relax}
\providecommand{\bibinfo}[2]{#2}
\providecommand{\BIBentrySTDinterwordspacing}{\spaceskip=0pt\relax}
\providecommand{\BIBentryALTinterwordstretchfactor}{4}
\providecommand{\BIBentryALTinterwordspacing}{\spaceskip=\fontdimen2\font plus
\BIBentryALTinterwordstretchfactor\fontdimen3\font minus \fontdimen4\font\relax}
\providecommand{\BIBforeignlanguage}[2]{{%
\expandafter\ifx\csname l@#1\endcsname\relax
\typeout{** WARNING: IEEEtran.bst: No hyphenation pattern has been}%
\typeout{** loaded for the language `#1'. Using the pattern for}%
\typeout{** the default language instead.}%
\else
\language=\csname l@#1\endcsname
\fi
#2}}
\providecommand{\BIBdecl}{\relax}
\BIBdecl

\bibitem{PopovVGSKW22_DIFFVC_ICLR}
\BIBentryALTinterwordspacing
V.~Popov, I.~Vovk, V.~Gogoryan, T.~Sadekova, M.~S. Kudinov, and J.~Wei, ``Diffusion-based voice conversion with fast maximum likelihood sampling scheme,'' in \emph{Proceedings of the International Conference on Learning Representations ({ICLR})}, 2022.
\BIBentrySTDinterwordspacing

\bibitem{LiuLRCZ22_DIFFSINGER_AIII}
\BIBentryALTinterwordspacing
J.~Liu, C.~Li, Y.~Ren, F.~Chen, and Z.~Zhao, ``DiffSinger: Singing voice synthesis via shallow diffusion mechanism,'' in \emph{Thirty-Sixth {AAAI} Conference on Artificial Intelligence, {AAAI} 2022}. {AAAI} Press, 2022, pp. 11\,020--11\,028.
\BIBentrySTDinterwordspacing

\bibitem{DaiLVG2024EXPRESSIVESINGER_ACMMULTIMEDIA}
\BIBentryALTinterwordspacing
S.~Dai, M.-Y. Liu, R.~Valle, and S.~Gururani, ``ExpressiveSinger: Multilingual and multi-style score-based singing voice synthesis with expressive performance control,'' in \emph{ACM Multimedia 2024}.
\BIBentrySTDinterwordspacing

\bibitem{DaiWDJ2025_Everyone_ICASSP}
\BIBentryALTinterwordspacing
S.~Dai, Y.~Wang, R.~B. Dannenberg, and Z.~Jin, ``Everyone-Can-Sing: Zero-shot singing voice synthesis and conversion with speech reference,'' in \emph{Proceedings of the {IEEE} International Conference on Acoustics, Speech, and Signal Processing ({ICASSP})}, 2025, pp. 1--5.
\BIBentrySTDinterwordspacing

\bibitem{HawthorneSRZGME22_SynthesisDiffusion_ISMIR}
\BIBentryALTinterwordspacing
C.~Hawthorne, I.~Simon, A.~Roberts, N.~Zeghidour, J.~Gardner, E.~Manilow, and J.~H. Engel, ``Multi-instrument music synthesis with spectrogram diffusion,'' in \emph{Proceedings of the International Society for Music Information Retrieval Conference ({ISMIR})}, 2022, pp. 598--607.
\BIBentrySTDinterwordspacing

\bibitem{MamanZMB24_PerformanceConditioning_ICASSP}
B.~Maman, J.~Zeitler, M.~M{\"u}ller, and A.~H. Bermano, ``Performance conditioning for diffusion-based multi-instrument music synthesis,'' in \emph{Proceedings of the {IEEE} International Conference on Acoustics, Speech, and Signal Processing ({ICASSP})}, 2024, pp. 5045--5049.

\bibitem{MamanZMB2024_MultiAspect_TASLP}
B.~Maman, J.~Zeitler, M.~Müller, and A.~H. Bermano, ``Multi-aspect conditioning for diffusion-based music synthesis: Enhancing realism and acoustic control,'' \emph{IEEE/ACM Trans. on Audio, Speech, and Lang. Process.}, pp. 1--14, 2024.

\bibitem{KameokaKTH2018_Stargan_SLT}
H.~Kameoka, T.~Kaneko, K.~Tanaka, and N.~Hojo, ``StarGAN-VC: Non-parallel many-to-many voice conversion using star generative adversarial networks,'' in \emph{IEEE Spoken Language Technology Workshop (SLT)}, 2018, pp. 266--273.

\bibitem{SismanYKL21_VCOVERVIEW_TASLP}
\BIBentryALTinterwordspacing
B.~Sisman, J.~Yamagishi, S.~King, and H.~Li, ``An overview of voice conversion and its challenges: From statistical modeling to deep learning,'' \emph{{IEEE} {ACM} Trans. Audio Speech Lang. Process.}, vol.~29, pp. 132--157, 2021.
\BIBentrySTDinterwordspacing

\bibitem{MehtaTBSH2024_Matcha_ICASSP}
S.~Mehta, R.~Tu, J.~Beskow, {\'E}.~Sz{\'e}kely, and G.~E. Henter, ``Matcha-TTS: A fast TTS architecture with conditional flow matching,'' in \emph{Proceedings of the {IEEE} International Conference on Acoustics, Speech, and Signal Processing ({ICASSP})}, 2024, pp. 11\,341--11\,345.

\bibitem{PiaSME2025_Flowmac_ICASSP}
N.~Pia, M.~Strauss, M.~Multrus, and B.~Edler, ``Flow{M}ac: Conditional flow matching for audio coding at low bit rates,'' in \emph{IEEE International Conference on Acoustics, Speech and Signal Processing (ICASSP), 2025}, pp. 1--5.

\bibitem{KalitaDSZHP24_PADVC_IWAENC}
A.~K. Kalita, C.~Dittmar, P.~Sani, F.~Zalkow, E.~A.~P. Habets, and R.~Patra, ``{PAD-VC}: {A} prosody-aware decoder for any-to-few voice conversion,'' in \emph{Proceedings of the International Workshop on Acoustic Signal Enhancement ({IWAENC})}, 2024, pp. 389--393.

\bibitem{ChurchwellMP2024_HIGH_ICASSPW}
C.~Churchwell, M.~Morrison, and B.~Pardo, ``High-fidelity neural phonetic posteriorgrams,'' in \emph{Proceedings of the {IEEE} International Conference on Acoustics, Speech, and Signal Processing ({ICASSP})}, 2024, pp. 823--827.

\bibitem{DaiWCHD23_SINGSTYLE111_ISMIR}
S.~Dai, Y.~Wu, S.~Chen, R.~Huang, and R.~B. Dannenberg, ``{S}ing{S}tyle111: {A} multilingual singing dataset with style transfer,'' in \emph{Proceedings of the International Society for Music Information Retrieval Conference ({ISMIR})}, 2023, pp. 765--773.

\bibitem{ShenJTLLHQZB24_NaturalSpeech_ICLR}
\BIBentryALTinterwordspacing
K.~Shen, Z.~Ju, X.~Tan, E.~Liu, Y.~Leng, L.~He, T.~Qin, S.~Zhao, and J.~Bian, ``NaturalSpeech 2: Latent diffusion models are natural and zero-shot speech and singing synthesizers,'' in \emph{Proceedings of the International Conference on Learning Representations ({ICLR})}, 2024.
\BIBentrySTDinterwordspacing

\bibitem{Copet2023_Simple_NEURIPS}
J.~Copet, F.~Kreuk, I.~Gat, T.~Remez, D.~Kant, G.~Synnaeve, Y.~Adi, and A.~D{\'e}fossez, ``Simple and controllable music generation,'' \emph{Advances in Neural Information Processing Systems}, vol.~36, pp. 47\,704--47\,720, 2023.

\bibitem{LeePGCY23_BIGVGAN_ICLR}
S.~Lee, W.~Ping, B.~Ginsburg, B.~Catanzaro, and S.~Yoon, ``{BigVGAN}: {A} universal neural vocoder with large-scale training,'' in \emph{Proceedings of the International Conference on Learning Representations ({ICLR})}, 2023.

\bibitem{PerezSVDC18_FiLM_AAAI}
\BIBentryALTinterwordspacing
E.~Perez, F.~Strub, H.~de~Vries, V.~Dumoulin, and A.~C. Courville, ``FiLM: Visual reasoning with a general conditioning layer,'' in \emph{Proceedings of the {AAAI} Conference on Artificial Intelligence}, 2018, pp. 3942--3951.
\BIBentrySTDinterwordspacing

\bibitem{HoS22_ClassifierFreeGuidance_arXiv}
\BIBentryALTinterwordspacing
J.~Ho and T.~Salimans, ``Classifier-free diffusion guidance,'' \emph{arXiv}, vol. abs/2207.12598, 2022.
\BIBentrySTDinterwordspacing

\bibitem{KimSLB18_CREPE_ICASSP}
J.~W. Kim, J.~Salamon, P.~Li, and J.~P. Bello, ``{CREPE}: {A} convolutional representation for pitch estimation,'' in \emph{Proceedings of the {IEEE} International Conference on Acoustics, Speech and Signal Processing ({ICASSP})}, 2018, pp. 161--165.

\bibitem{BaevskiZMA20_WAV2VEC2_NEURIPS}
A.~Baevski, Y.~Zhou, A.~Mohamed, and M.~Auli, ``{wav2vec 2.0}: {A} framework for self-supervised learning of speech representations,'' in \emph{Advances in Neural Information Processing Systems ({NeurIPS})}, 2020.

\bibitem{BabuWTLXGSPSPBC22_XLSR_INTERSPEECH}
A.~Babu, C.~Wang, A.~Tjandra, K.~Lakhotia, Q.~Xu, N.~Goyal, K.~Singh, P.~von Platen, Y.~Saraf, J.~Pino, A.~Baevski, A.~Conneau, and M.~Auli, ``{XLS-R}: {S}elf-supervised cross-lingual speech representation learning at scale,'' in \emph{Proceedings of the Annual Conference of the International Speech Communication Association (Interspeech)}, 2022, pp. 2278--2282.

\bibitem{RouardMD23_HTDemucs_ICASSP}
S.~Rouard, F.~Massa, and A.~D{\'e}fossez, ``Hybrid transformers for music source separation,'' in \emph{Proceedings of the {IEEE} International Conference on Acoustics, Speech, and Signal Processing ({ICASSP})}, 2023.

\bibitem{HawthorneESRSRE18_OnsetsFrames_ISMIR}
C.~Hawthorne, E.~Elsen, J.~Song, A.~Roberts, I.~Simon, C.~Raffel, J.~H. Engel, S.~Oore, and D.~Eck, ``Onsets and Frames: {D}ual-objective piano transcription,'' in \emph{Proceedings of the International Society for Music Information Retrieval Conference, ({ISMIR})}, 2018, pp. 50--57.

\bibitem{HawthorneSRSHDE19_MAESTRO_ICLR}
\BIBentryALTinterwordspacing
C.~Hawthorne, A.~Stasyuk, A.~Roberts, I.~Simon, C.~A. Huang, S.~Dieleman, E.~Elsen, J.~H. Engel, and D.~Eck, ``Enabling factorized piano music modeling and generation with the {MAESTRO} dataset,'' in \emph{Proceedings of the International Conference on Learning Representations ({ICLR})}, 2019.
\BIBentrySTDinterwordspacing

\bibitem{MamanBermano22_UnalignedAMT_ICML}
B.~Maman and A.~H. Bermano, ``Unaligned supervision for automatic music transcription in the wild,'' in \emph{Proceedings of the International Conference on Machine Learning ({ICML})}, 2022, pp. 14\,918--14\,934.

\bibitem{YaffeMMB25_CountNotes_ISMIR}
J.~Yaffe, B.~Maman, M.~M{\"u}ller, and A.~Bermano, ``Count The Notes: {H}istogram-based supervision for automatic music transcription,'' in \emph{Proceedings of the International Society for Music Information Retrieval Conference ({ISMIR})}, Daejeon, South Korea, 2025, pp. 469--476.

\bibitem{ShorJMLTQTSEH20_NonSemanticSpeechRepresentation_Interspeech}
\BIBentryALTinterwordspacing
J.~Shor, A.~Jansen, R.~Maor, O.~Lang, O.~Tuval, F.~de~Chaumont~Quitry, M.~Tagliasacchi, I.~Shavitt, D.~Emanuel, and Y.~Haviv, ``Towards learning a universal non-semantic representation of speech,'' in \emph{Proc. of the Annual Conference of the International Speech Communication Association (Interspeech)}, 2020, pp. 140--144.
\BIBentrySTDinterwordspacing

\bibitem{ZalkowSKHPD25_LowResourceGenerativePostprocessing_INTERSPEECH}
F.~Zalkow, P.~Sani, K.~K. Lakshminarayana, E.~A.~P. Habets, N.~Pia, and C.~Dittmar, ``Bridging the training–inference gap in {TTS}: {T}raining strategies for robust generative postprocessing for low-resource speakers,'' in \emph{Proceedings of the Conference of the International Speech Communication Association (INTERSPEECH)}, 2025.

\bibitem{WeissZAMKVG21_WinterreiseDataset_ACM-JOCCH}
C.~Wei{\ss}, F.~Zalkow, V.~Arifi-M{\"u}ller, M.~M{\"u}ller, H.~V. Koops, A.~Volk, and H.~Grohganz, ``{S}chubert {W}interreise dataset: {A} multimodal scenario for music analysis,'' \emph{{ACM} Journal on Computing and Cultural Heritage ({JOCCH})}, vol.~14, no.~2, pp. 25:1--18, 2021.

\bibitem{KilgourZRS19_FAD_Interspeech}
\BIBentryALTinterwordspacing
K.~Kilgour, M.~Zuluaga, D.~Roblek, and M.~Sharifi, ``Fr{\'{e}}chet Audio Distance: {A} reference-free metric for evaluating music enhancement algorithms,'' in \emph{Proceedings of the Annual Conference of the International Speech Communication Association (Interspeech)}, 2019, pp. 2350--2354.
\BIBentrySTDinterwordspacing

\bibitem{ITU15}
{I}nternational~{T}elecommunications {U}nion, ``{ITU-R} {R}ec. {BS}.1534-3: Method for the subjective assessment of intermediate quality levels of coding systems,'' 2015.

\bibitem{SchoefflerBSRWEH18_webMUSHRA_JORS}
M.~Schoeffler, S.~Bartoschek, F.-R. St{\"o}ter, M.~Roess, S.~Westphal, B.~Edler, and J.~Herre, ``{webMUSHRA}—a comprehensive framework for web-based listening tests,'' \emph{Journal of Open Research Software}, vol.~6, no.~1, 2018.

\bibitem{PolinerEEGSO07_MelodyTranscription_TASLP}
G.~E. Poliner, D.~P. Ellis, A.~F. Ehmann, E.~G{\'o}mez, S.~Streich, and B.~Ong, ``Melody transcription from music audio: Approaches and evaluation,'' \emph{{IEEE} Trans. on Audio, Speech, and Lang. Process.}, vol.~15, no.~4, pp. 1247--1256, 2007.

\bibitem{SalamonG12_MelodyExtraction_TASLP}
J.~Salamon and E.~G{\'o}mez, ``Melody extraction from polyphonic music signals using pitch contour characteristics,'' \emph{IEEE Transactions on Audio, Speech, and Language Processing}, vol.~20, no.~6, pp. 1759--1770, 2012.

\bibitem{RaffelMHSNLE14_MirEval_ISMIR}
C.~Raffel, B.~McFee, E.~J. Humphrey, J.~Salamon, O.~Nieto, D.~Liang, and D.~P.~W. Ellis, ``{MIR{\_}EVAL}: {A} transparent implementation of common {MIR} metrics,'' in \emph{Proceedings of the International Society for Music Information Retrieval Conference ({ISMIR})}, 2014, pp. 367--372.

\bibitem{FugledeT2004_Jensen_ISIT}
B.~Fuglede and F.~Topsoe, ``Jensen--Shannon divergence and Hilbert space embedding,'' in \emph{IEEE International symposium on Information theory ({ISIT})}, 2004, p.~31.

\bibitem{Vaserstein1969_Markov}
L.~N. Vaserstein, ``Markov processes over denumerable products of spaces, describing large systems of automata,'' \emph{Problemy Peredachi Informatsii}, vol.~5, no.~3, pp. 64--72, 1969.

\end{thebibliography}
\end{document}